\pdfoutput=1

\documentclass[3p,times,procedia]{elsarticle}
\usepackage{nupha_ecrc}

\newcommand{\pp}{\ensuremath{\mathrm {p\kern-0.05em p}}}

\newcommand{\sqrtSnn}{\ensuremath{\sqrt{s_{\mathrm{NN}}}}}

\newcommand{\pt}{\ensuremath{p_{\mathrm{T}}}}

\newcommand{\MeVc}{\ensuremath{\mathrm{MeV}\kern-0.05em/\kern-0.02em c}}

\newcommand{\GeVc}{\ensuremath{\mathrm{GeV}\kern-0.05em/\kern-0.02em c}}

\newcommand{\GeVcSq}{\ensuremath{\mathrm{GeV}\kern-0.05em/\kern-0.02em c^2}}

\newcommand{\jpsi}{\ensuremath{{\rm J}\kern-0.02em/\kern-0.05em\psi}}

\newcommand{\dpt}{\ensuremath{\delta \kern-0.15em p_{\mathrm{T}}}}

\newcommand{\rpn}{\ensuremath{\Psi_n}}

\volume{00}

\firstpage{1}

\journalname{Nuclear Physics A}

\runauth{R. A. Bertens}

\jid{nupha}

\jnltitlelogo{Nuclear Physics A}

\usepackage{amssymb}

\usepackage[figuresright]{rotating}

\begin{document}

\begin{frontmatter}



    \dochead{XXVIth International Conference on Ultrarelativistic Nucleus-Nucleus Collisions\\ (Quark Matter 2017)}

    \title{Anisotropic flow of inclusive and identified particles in Pb--Pb collisions at $\sqrt{s_{\rm NN}} = 5.02$ TeV}


    \author{R. A. Bertens (for the ALICE collaboration)}
    \ead{redmer.alexander.bertens@cern.ch}

    \address{University of Knoxville (Tennessee USA), CERN}

    \begin{abstract}
        Anisotropic flow measurements constrain the shear $(\eta/s)$ and bulk ($\zeta/s$) viscosity of the quark-gluon plasma created in heavy-ion collisions, as well as give insight into the initial state of such collisions and hadronization mechanisms. In these proceedings, elliptic ($v_2$) and higher harmonic ($v_3, v_4$) flow coefficients of $\pi^{\pm}$, $K^{\pm}$, p$(\overline{\rm{p}})$ and the $\phi$-meson, measured in Pb–-Pb collisions at the highest-ever center-of-mass energy of $\sqrt{s_{\rm NN}}$ = 5.02 TeV, are presented. 
    \end{abstract}


\end{frontmatter}


\section{Introduction}
Heavy-ion collision experiments are used to study the properties of the quark-gluon plasma (QGP), a state of deconfined quarks and gluons created at high baryon densities or temperatures. Particles produced in collisions are boosted collectively by a common velocity field that is induced by the rapid expansion of the system. Spatial anisotropies resulting from the elliptic overlap region of the colliding nuclei and the initial inhomogeneities of the system density are transformed, through multiple interactions between the produced particles, into an anisotropy in momentum space of the produced particles. The efficiency of this process depends on e.g. the shear ($\eta/s$) and bulk ($\zeta/s$) viscosity of the created matter, and the lifetime of the system. 

Anisotropy in particle production can be quantified by a Fourier analysis of the azimuthal distribution relative to the system's symmetry plane angles \rpn{}, characterized by harmonic coefficients $v_n$ \cite{olli} 
\begin{equation}\label{eq:flowdud}
    \frac{\mathrm{d}N}{\mathrm{d}\left(\varphi - \rpn\right)} \propto 1 + \sum_{n=1}^{\infty} 2 v_n \cos\left(n\left[\varphi - \rpn\right]\right),
\end{equation}
where $\varphi$ is the azimuthal angle of the produced particles.

Flow coefficients $v_n$ are, in addition to being a probe for $\eta/s$ and $\zeta/s$, sensitive to the initial state of the system,  freeze-out conditions and hadronization mechanisms.

\section{Data analysis}
The data used for this work were recorded in 2015 at a center of mass energy per nucleon of \sqrtSnn{} = 5.02 TeV with the ALICE detector \cite{performance} and comprise approximately 6$\cdotp10^7$ collisions with a vertex within $\pm$10 cm of the nominal interaction point and collision centrality between 5-60\%. Charged particle tracks are reconstructed using the Inner Tracking System (ITS) and Time Projection Chamber (TPC) at $\vert y \vert <$ 0.5 for identified particles or $\vert \eta \vert < 0.8$ for unidentified particles. Centrality determination, as well as reconstruction of the \textbf{Q}$_n$ vectors (see Eq.~\ref{eq:mth_sp}), is performed with V0 detectors, located at 2.8 $< \eta <$ 5.1 and -3.7 $< \eta <$ -1.7. 

Coefficients $v_n$ are measured using the scalar product method \cite{Voloshin:2008dg}, written as
\begin{equation}
    v_{n}\{{\rm SP, V0C}\} = \langle \langle {\bf u}_n \cdotp {\bf Q}_n^{\rm V0C *} \rangle \rangle \Bigg/ \sqrt{ \frac{\langle {\bf Q}_n^{\rm V0C}\cdotp {\bf Q}_n^{\rm TPC *} \rangle \langle  {\bf Q}_n^{\rm V0C}\cdotp {\bf Q}_n^{\rm V0A *} \rangle} { \langle  {\bf Q}_n^{\rm TPC} \cdotp{\bf Q}_n^{\rm V0A *} \rangle } } 
    \label{eq:mth_sp}
\end{equation}
where ${\bf u}_n=\exp(in\varphi)$ is the unit vector of a single particle with azimuthal angle $\varphi$. Flow vectors \textbf{Q}$_n$ = $\sum_j \exp(i n \varphi_j)$, where the sum runs over all $j$ tracks and $^*$ denotes the complex conjugate, are measured in the TPC or V0 detectors. Brackets $\langle \dots \rangle$ indicate an all-event average; the double brackets in the numerator of Eq.~\ref{eq:mth_sp} mean that prior to the all-event average, an average over all tracks within the single event is taken. The large (pseudo-)rapidity gap between \textbf{u}$_n$ and \textbf{Q}$^{\rm V0C}_n$ reduces sensitivity to short-range correlations that are unrelated to the initial geometry, commonly referred to as \emph{non-flow}. 

Particle identification is performed using ionization energy loss measured in the TPC, combined with the arrival time of particles in the Time of Flight (TOF) detector. The $\phi$-meson is reconstructed in the $\phi \rightarrow K^+ K^-$ channel, using the analysis method outlined in \cite{pidv2}; its $v_2$ is determined using Eq.~\ref{eq:mth_sp}. 

\section{Results}
\begin{figure}
    \begin{center}
        \includegraphics[width=\textwidth]{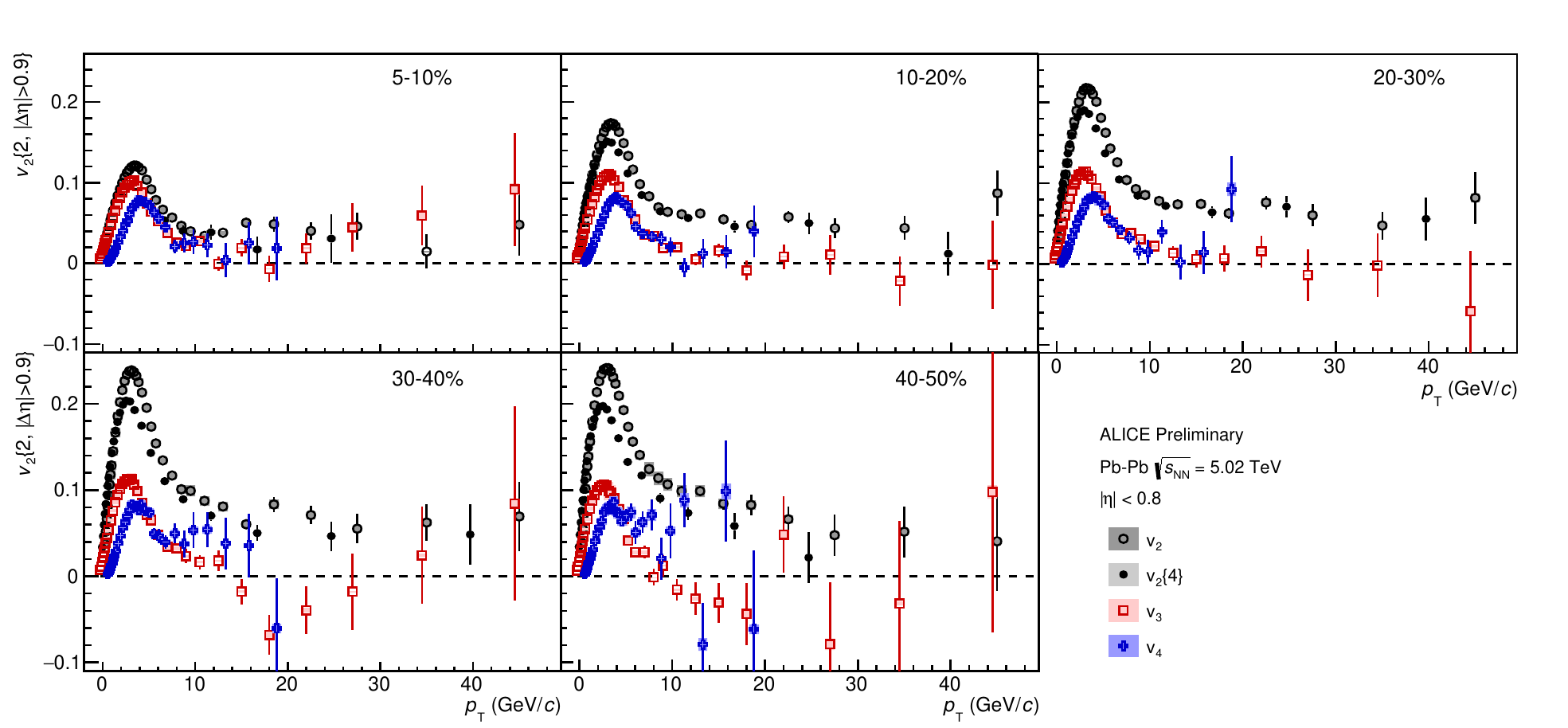}
        \caption{Flow coefficients $v_2$, $v_3$ and $v_4$ of unidentified charged particles as function of \pt{} for various centrality classes. In addition to the scalar product method, the 4-particle $q$-cumulant \cite{qcumulants} estimate of $v_2$, $v_2\{4\}$, is shown as well. Statistical uncertainties are shown as bars and systematic uncertainties as boxes.}
        \label{fig:drawHistVnPtAll}
    \end{center}
\end{figure}
Figure~\ref{fig:drawHistVnPtAll} shows \pt{}-differential $v_2$, $v_3$ and $v_4$ of unidentified charged particles. For the presented centrality classes, $v_2 > v_3 > v_4$ for \pt{} $<$ 5 \GeVc{}. The observed trends at low and intermediate \pt{} ($<$ 7 \GeVc{}) are characteristic for the hydrodynamic expansion of the medium. The non-zero $v_n$ at high \pt{} is attributed to path-length dependent in-medium energy loss of highly energetic partons. 

\begin{figure}
    \begin{center}
        \includegraphics[width=\textwidth]{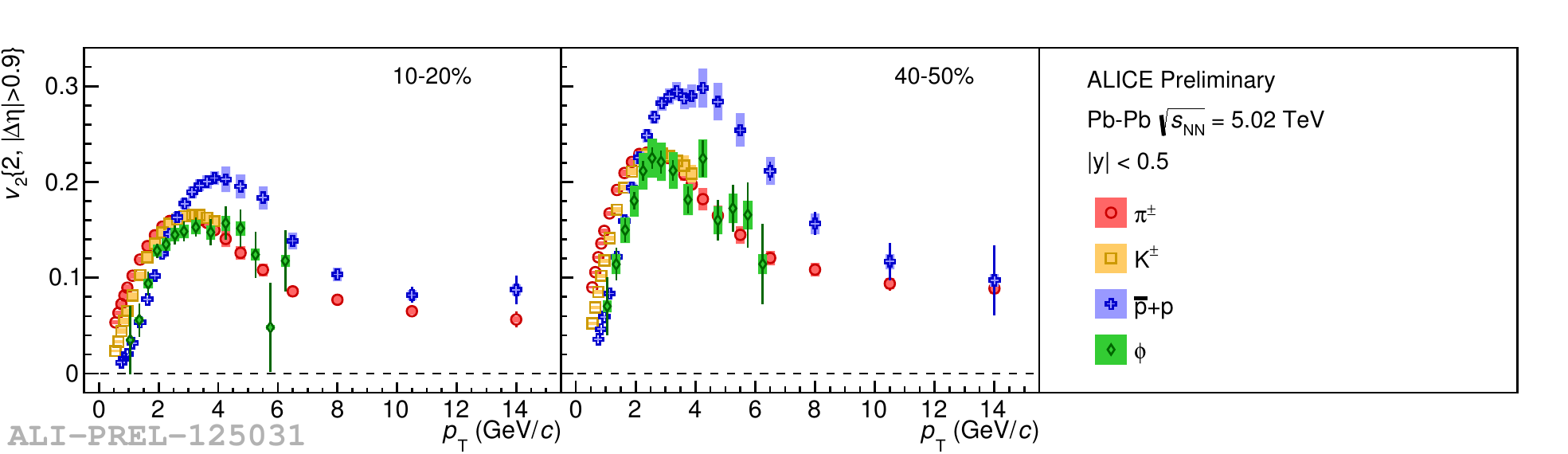}
        \includegraphics[width=\textwidth]{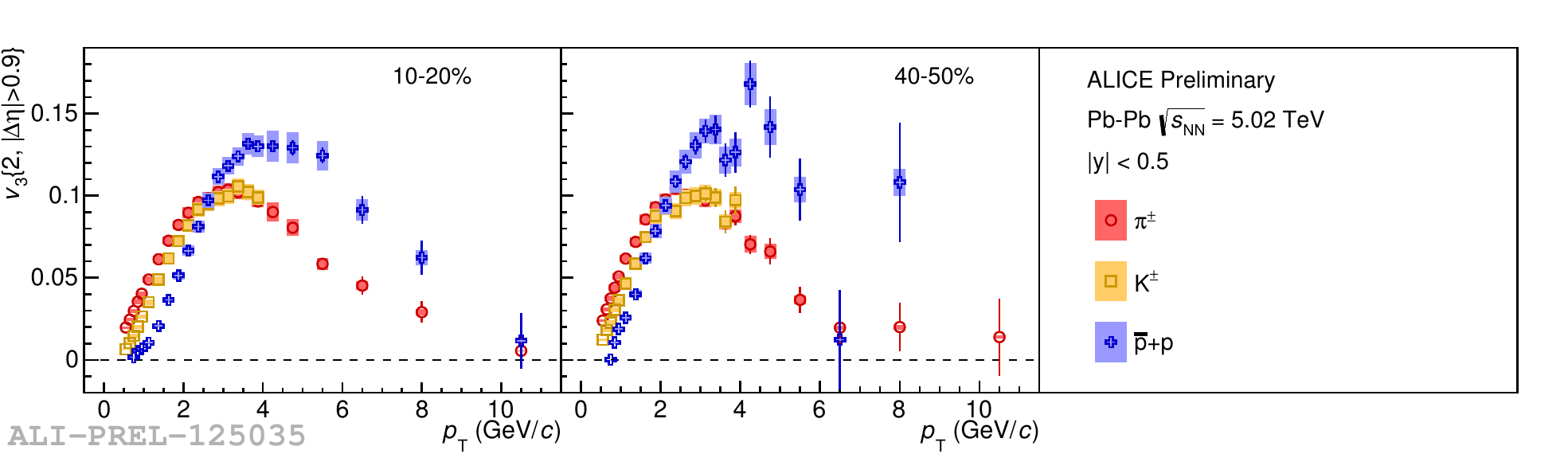}
        \includegraphics[width=\textwidth]{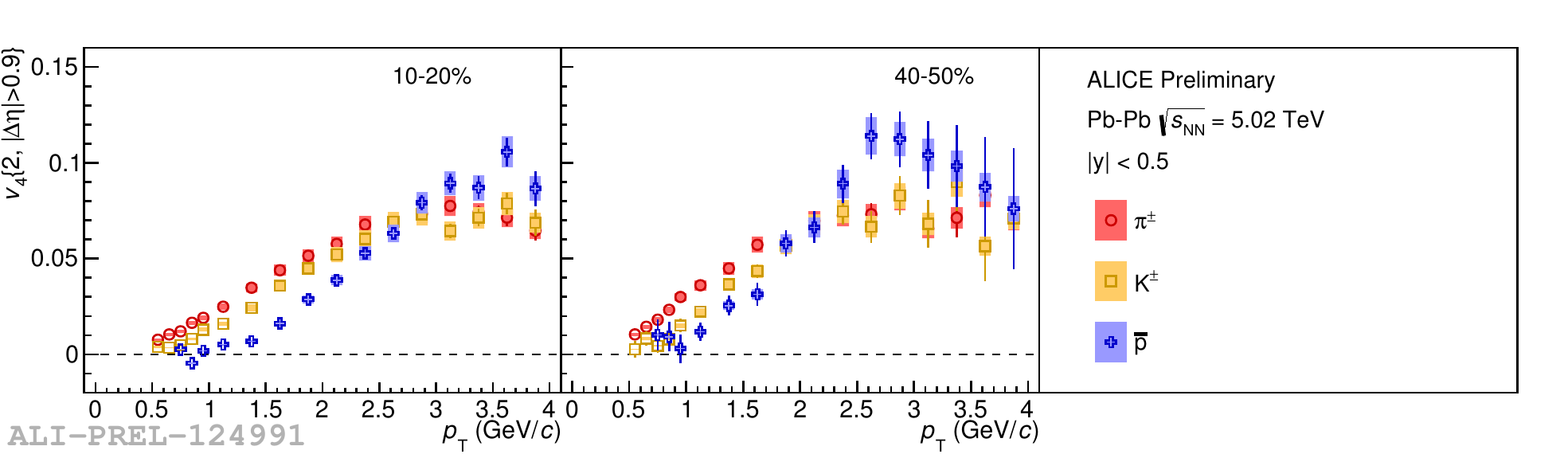}
        \caption{Flow coefficient $v_2$ (top), $v_3$ (middle) and $v_4$ (bottom) of $\pi^{\pm}, K^{\pm}$, p($\overline{\rm p})$ and the $\phi$-meson for 10-20\% (left) and 40-50\% (right) collision centrality as function of \pt{}. Statistical uncertainties are shown as bars and systematic uncertainties as boxes.}
            \label{fig:v2_pid}
        \end{center}
    \end{figure}

The top panel of Fig.~\ref{fig:v2_pid} shows \pt{}-differential $v_2$ of $\pi^{\pm}, K^{\pm}$, p$(\overline{\rm{p}})$ and the $\phi$-meson for 10-20\% (left) and 40-50\% (right) collision centrality (these two centrality intervals are used for all subsequent figures). For \pt{}~$<$~2~GeV/$c$, $v_2$ of the different species is mass-ordered, which is indicative of strong radial flow. For 3~$<$~\pt{}~$<$~8~\GeVc{}, particles are grouped according to their valence quark content, which supports the hypothesis of particle production via quark coalescence \cite{dud2}. Particle type scaling and mass ordering is most directly tested by $\phi$-meson $v_2$, as the $\phi$ is a meson with a mass close to proton mass. Figure~\ref{fig:v2_pid} demonstrates that $\phi$-meson $v_2$ follows proton $v_2$ at low \pt{}, but pion $v_2$ at intermediate \pt{}. Lastly it should be noted that p($\overline{\rm p}$) $v_2$ is larger than $\pi^{\pm}$ $v_2$ for $3 \lesssim \pt{} \lesssim$ 10 \GeVc{}, after which the $v_2$ converge, which suggests that partonic energy loss is flavor independent at high transverse momenta. 

    Higher harmonic flow coefficients ($n > 2$) are generated by inhomogeneities in the initial nucleon distribution and are thought to be more sensitive to transport coefficients than $v_2$ \cite{dud3}. The middle and lower panels of Fig.~\ref{fig:v2_pid} show that non-zero $v_3$ is observed for $\pi^{\pm}, K^{\pm}$, p($\overline{\rm p})$ up to \pt{} $\approx$ 8 \GeVc{}. Statistical precision limits the range of the $v_4$ measurement to \pt{} $<$ 4 \GeVc{}; $v_4$ is non-zero though in the entire measured range. Both $v_3$ and $v_4$ show a clear mass ordering at low \pt{}, and analogous to the trend of $v_2$, p($\overline{\rm p}$) $v_3$ is larger than $\pi^{\pm}$ $v_3$ up to \pt{} = 10 \GeVc{}. The crossing of the meson and baryon trends at $\pt{} \approx$ 2.5 \GeVc{} is reminiscent of the behavior observed for $v_2$ as well. Overall, the $v_n$ values are qualitatively similar to those observed at a collision energy of \sqrtSnn{} = 2.76 TeV \cite{pidv2,naghmeh}.

    \begin{figure}
        \begin{center}
            \includegraphics[width=\textwidth]{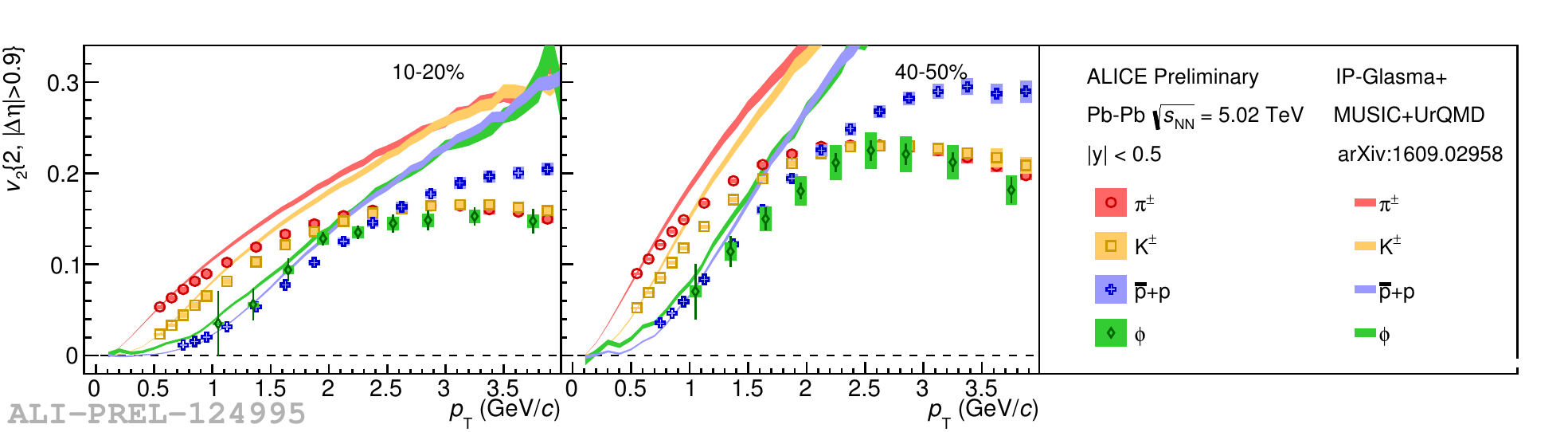}
            \caption{Flow coefficient $v_2$ of $\pi^{\pm}, K^{\pm},$ p$(\overline{\rm p})$ and the $\phi$-meson for 10-20\% (left) and 40-50\% (right) collision centrality compared to predictions from relativistic hydrodynamic calculations \cite{dud4}. Statistical uncertainties are shown as bars and systematic uncertainties as boxes.}
                \label{fig:v2_pid_hydro}
            \end{center}
        \end{figure}

        To test the validity of the hydrodynamic description of the QGP, $v_n$ are compared to model predictions from \cite{dud4} in Fig.~\ref{fig:v2_pid_hydro}. The model uses an IP-Glasma initial state and a viscous hydrodynamic medium evolution ($\eta/s$ = 0.095 and a temperature-dependent $\zeta/s$) which is coupled to a hadronic cascade procedure for hadronization. Interestingly, mass ordering is broken ($\phi$-meson $v_2 >$  p($\overline{\rm p}$) $v_2$) in the calculations. The predictions show good agreement with the data for $\pt{} <$ 1 \GeVc{} in central collisions, but overestimate $v_2$ already at lower momenta for more peripheral collisions. Similar behavior is found for $v_3$ and $v_4$ (not shown here). 

        To test the hypothesis of particle production via quark coalescence, the axes of Fig.~\ref{fig:v2_pid} can be scaled by the number of constituent quarks, independently for each species. Such a scaling (not shown) shows that from \pt{}$/n_{\rm q} >$ 1.5 \GeVc{} particles group approximately according to their type (baryon, meson), similar behavior is observed for $v_3$ and $v_4$. It is stressed that the observed scaling only holds approximately, as was also observed elsewhere \cite{pidv2}. 

            \section{Summary}
            Flow harmonics $v_2, v_3$ and $v_4$ of unidentified and identified particles have been measured at \sqrtSnn{} = 5.02 TeV Pb--Pb collisions. Mass ordering is observed for \pt{} $<$ 2 \GeVc{}, as well as approximate particle type scaling for \pt{} $>$ 2.5 \GeVc{}. The flow coefficient $v_2$ of unidentified particles is non-zero up to high \pt{}, and p($\overline{\rm p}$) $v_2$, $v_3$ are larger than $\pi^{\pm}$ $v_2$, $v_3$ up to \pt{} = 10 \GeVc{}. The unprecedented precision of these new measurements will put strong constraints on model calculations and furthers the understanding of the hydrodynamic behavior of the QGP, as well as its initial state, and freeze-out conditions. 

            \label{}





            \bibliographystyle{elsarticle-num}
            \bibliography{references.bib}







            \end{document}